\documentclass[prl,twocolumn,nofootinbib,showpacs,superscriptaddress,preprintnumbers,amsmath,amssymb]{revtex4-2}
\usepackage[colorlinks=true, pdfstartview=FitV, linkcolor=red, citecolor=blue, urlcolor=black, pdftitle={},pdfauthor={Tomoya Hayata},pdfsubject={}, pdfkeywords={}]{hyperref}
\usepackage{graphicx}
\usepackage{times,setspace,bm}
\usepackage{latexsym,amssymb,amsmath,mathrsfs,mathtools}
\usepackage{color}
\graphicspath{{../figure/}}

\makeatletter
\providecommand \@ifxundefined [1]{%
 \@ifx{#1\undefined}
}%
\providecommand \@ifnum [1]{%
 \ifnum #1\expandafter \@firstoftwo
 \else \expandafter \@secondoftwo
 \fi
}%
\providecommand \@ifx [1]{%
 \ifx #1\expandafter \@firstoftwo
 \else \expandafter \@secondoftwo
 \fi
}%
\providecommand \bibnamefont  [1]{#1}%
\providecommand \bibfnamefont [1]{#1}%
\providecommand \citenamefont [1]{#1}%
\providecommand \href@noop [0]{\@secondoftwo}%
\providecommand \href [0]{\begingroup \@sanitize@url \@href}%
\providecommand \@href[1]{\@@startlink{#1}\@@href}%
\providecommand \@@href[1]{\endgroup#1\@@endlink}%
\providecommand \@sanitize@url [0]{\catcode `\\12\catcode `\$12\catcode
  `\&12\catcode `\#12\catcode `\^12\catcode `\_12\catcode `\%12\relax}%
\providecommand \@@startlink[1]{}%
\providecommand \@@endlink[0]{}%
\providecommand \url  [0]{\begingroup\@sanitize@url \@url }%
\providecommand \@url [1]{\endgroup\@href {#1}{\urlprefix }}%
\providecommand \urlprefix  [0]{URL }%
\providecommand \Eprint [0]{\href }%
\providecommand \selectlanguage [0]{\@gobble}%
\providecommand \bibinfo  [0]{\@secondoftwo}%
\providecommand \bibfield  [0]{\@secondoftwo}%
\providecommand \BibitemOpen [0]{}%
\providecommand \EOS [0]{\spacefactor3000\relax}%
\providecommand \BibitemShut  [1]{\csname bibitem#1\endcsname}%
\let\auto@bib@innerbib\@empty

\newcommand{\cL}{{\cal L}}

\newcommand{\up}{\uparrow}
\newcommand{\down}{\downarrow}
\newcommand{\be}{\begin{equation}}      
\newcommand{\ee}{\end{equation}}      
\newcommand{\bea}{\begin{eqnarray}}      
\newcommand{\eea}{\end{eqnarray}}
\newcommand{\tr}{\mathrm{tr}}

\begin{document}
\title{Lindblad evolution without the sign problem}

\author{Tomoya Hayata}
\affiliation{
Department of Physics, Keio University, 4-1-1 Hiyoshi, Kanagawa 223-8521, Japan
}

\date{\today}

\begin{abstract}
Quantum Monte Carlo is one of the most powerful numerical tools for studying nonpeturbative properties of quantum many-body systems.
However, its application to real-time problems is limited since the complex and highly-oscillating path-integral weight of the real-time evolution harms the important sampling. 
In this Letter, we show that some real-time problems in open fermion systems can be simulated using the quantum Monte Carlo.
To this end, we prescribe a mapping between a real-time problem in open quantum systems and a statistical problem in non-Hermitian quantum systems; for some cases, the latter can be solved without suffering from the complex measure problem. 
To explain our idea and demonstrate how it works, we compute the real-time evolution of fidelities in open fermion systems under dissipation. 

\end{abstract}

\maketitle

\paragraph{Introduction.} 
Understanding the real-time dynamics of quantum many-body systems under dissipations due to environments or measurements has been one of the most important and challenging problems in modern physics.
Intriguing problems drawing attention of researchers are e.g., the dissipation-induced phase transition~\cite{Diehl_2008,Sieberer_2016}, entanglement phase transition~\cite{PhysRevB.99.224307,PhysRevX.9.031009,PhysRevB.100.134306}, information scrambling under decoherence~\cite{PhysRevX.9.011006,Landsman_2019,Hayata:2021kcp}, as well as how quantum computing works in a noisy near-term quantum device~\cite{Preskill2018quantumcomputingin}.
Even limiting the scope of our interests to the Markovian cases, where the time evolution of an open quantum system is described by the quantum master equation in the Lindblad form~\cite{doi:10.1063/1.522979,Lindblad:1975ef}, the efficient numerical methods are still under development (see e.g., Ref.~\cite{RevModPhys.93.015008} for a review).

The most promising candidate for accurate and large-scale numerical simulations would be a family of quantum Monte Carlo methods if they are applicable.
However real-time problems suffer from the severe sign problem than statistical problems (see e.g. Ref.~\cite{PhysRevLett.117.081602}).
We will explain that in more detail.
In a statistical problem, we compute the partition function based on the importance sampling of  the imaginary-time (Suzuki-Trotter) evolution.
When the imaginary-time evolution becomes sign alternating, the importance sampling is no longer applicable; This is the notorious sign problem~\cite{PhysRevLett.94.170201}. 
Similarly, in a real-time problem, we need to compute the real-time evolution based on the importance sampling.
However, the path-integral weight has now a complex and highly oscillating phase, which strongly harms the importance sampling.

Even though the conventional Monte Carlo approach suffers from the severe sign problem, 
in this Letter, we show that the real-time evolution of some observable in open fermion systems can be computed by the quantum Monte Carlo.
The key idea is the correspondence between the Lindblad evolution and imaginary-time evolution with a non-Hermitian Hamiltonian.
Based on it, we provide a prescription by mapping a real-time problem in open quantum systems to a statistical problem in non-Hermitian quantum systems.
Although the non-Hermiticity usually harms the importance sampling, we can avoid it if the non-Hermitian Hamiltonian has a special symmetry~\cite{Hayata:2021erf}.
This implies that if we can map a real-time problem to a statistical problem with such special non-Hermitian system, we can solve the real-time problem by the quantum Monte Carlo. 
For demonstration of our idea, we compute the real-time evolution of fidelities in open fermion systems by the determinant quantum Monte Carlo.

\paragraph{Quantum master equation.} We consider an open quantum system interacting with environments or under continuous measurements.
If the effects of environments are perturbative and approximated by the Markov process, the time evolution of a density matrix of the system $\rho$ is described by the quantum master equation in the Lindblad form~\cite{doi:10.1063/1.522979,Lindblad:1975ef}:
\be
\begin{split}
\frac{d\rho}{dt} &=-i\left(H\rho-\rho H\right)
\\
&+\sum_i\gamma_i\left( \Gamma_i \rho \Gamma_i^\dagger-\frac{1}{2}\Gamma_i^\dagger\Gamma_i\rho-\frac{1}{2}\rho\Gamma_i^\dagger\Gamma_i \right) .
\end{split}
\label{eq:lindblad}
\ee
Here, $t$ and $H$ are the time and Hamiltonian operator, respectively. 
The first term in the right-hand side of Eq.~\eqref{eq:lindblad} gives the unitary time-evolution.
On the other hand, the second term gives the non-unitary time-evolution due to dissipations;
$\Gamma_i$, and $\gamma_i>0$ are the $i$-th quantum jump operator and its strength, respectively.
The quantum jump operators act on a density matrix by a superposition as in the second term of Eq.~\eqref{eq:lindblad}; This superposition keeps the completely positive and trace-preserving properties of $\rho$ during the time evolution.

As is well known, Eq.~\eqref{eq:lindblad} can be rewritten in the Sch{\"o}dinger equation type-form, namely, the matrix and vector form (see e.g., Ref.~\cite{RevModPhys.93.015008}).
To this end, we map a density matrix to a wave function (vector) in the doubled Hilbert space:
\be
\rho=\sum_{i,j}\rho_{ij}|i\rangle\langle j| \rightarrow |\rho\rangle=\sum_{i,j}\rho_{ij}|i\rangle\otimes |j\rangle ,
\ee
where $|i\rangle$, and $|j\rangle$ are the basis of the original Hilbert space.
In this representation, Eq.~\eqref{eq:lindblad} is written as
\be
\frac{d|\rho\rangle}{dt}=\cL |\rho\rangle ,
\ee
with the Liouville operator
\be
\begin{split}
\cL  &= -iH\otimes \bm 1+i\bm 1\otimes H^T
 \\
&+\sum_i\gamma_i\left( \Gamma_i \otimes \Gamma_i^*-\frac{1}{2}\Gamma_i^\dagger\Gamma_i\otimes\bm1-\frac{1}{2}\bm1\otimes\Gamma_i^T\Gamma_i^* \right) .
\end{split}
\label{eq:Liouville}
\ee
In this representation, the time evolution is formally solved as $|\rho(t)\rangle=e^{\cL t}|\rho(0)\rangle$ with the initial state  $|\rho(0)\rangle$.
As we shall see below, the real-time evolution operator $e^{\cL t}$ is understood as the imaginary-time evolution operator with the non-Hermitian Hamiltonian, and the quantum Monte Carlo is adopted to evaluate the imaginary-time evolution.

\paragraph{Fidelities and quantum Monte Carlo.} Let us define two fidelities, which are computable using the quantum Monte Carlo.
One is a generalization of the Loschmidt echo to open quantum systems:
\be
\begin{split}
M_\rho(t) &=\langle \rho_{\gamma_i=0}(t)|\rho(t)\rangle
\\
&=\langle \rho(0)|e^{-\cL_{\gamma_i=0} t}e^{\cL t}|\rho(0)\rangle ,
\label{eq:echo_pro}
\end{split}
\ee
where $|\rho_{\gamma_i=0}(t)\rangle$, and $\cL_{\gamma_i=0}$ are the vectorized density matrix, and evolution operator without quantum jumps, i.e., all $\gamma_i=0$ in Eq.~\eqref{eq:Liouville}, and $|\rho_{\gamma_i=0}(t)\rangle$ obeys the Hamiltonian evolution. 
Regarding the quantum jumps as perturbations, this quantifies the deviation of trajectories from the Hamiltonian evolution due to dissipations.
The other is a relative purity:
\be
\begin{split}
P_\rho(t) &=\langle \rho(0)|\rho(t)\rangle
\\
&=\langle \rho(0)|e^{\cL t}|\rho(0)\rangle .
\label{eq:purity_pro}
\end{split}
\ee
This is an extension of the persistent probability in the unitary system, and used e.g.,  for studying quantum speed limits in open systems~\cite{PhysRevLett.110.050403}.
Here, to make a connection between real-time and statistical problems apparent, we consider average about initial states $\sum_{\rho(0)}M_\rho$ and  $\sum_{\rho(0)}P_\rho$.
As initial states, we consider all eigenvectors of the computational basis of the doubled Hilbert space ($|\rho(0)\rangle=|i\rangle\otimes |j\rangle$). 
Then, we can replace the inner product by the trace as 
\be
M(t) =\tr\; e^{-\cL_{\gamma_i=0} t}e^{\cL t} ,
\label{eq:echo}
\ee
and
\be
P(t) =\tr\; e^{\cL t} .
\label{eq:purity}
\ee

To demonstrate how the quantum Monte Carlo works, let us solve a concrete model.
We consider a spinless fermion in a two-dimensional square lattice with regular hopping terms, and impose the periodic boundary conditions.
We choose the number operator at each site $\bm x$ as quantum jump operator, and take $\gamma_i=\gamma$ for simplicity (It is generalizable to the inhomogeneous and even time-dependent couplings).
Then, the Liouville operator reads
\be
\cL = K+U ,
\label{eq:Liouville_fermion}
\ee
with
\be
\begin{split}
K =&-iw\sum_{\bm x,i} \left(c^\dagger_{\bm x} c_{\bm x+\hat{i}}+c^\dagger_{\bm x+\hat{i}} c_{\bm x}\right)
\\
&+iw\sum_{\bm x,i} \left(d^\dagger_{\bm x} d_{\bm x+\hat{i}}+d^\dagger_{\bm x+\hat{i}} d_{\bm x}\right) ,
 \end{split}
\label{eq:hopping}
 \ee
and
\be
U=\sum_{\bm x}\gamma  \left(c^\dagger_{\bm x} c_{\bm x}-\frac{1}{2}\right) \left( d^\dagger_{\bm x} d_{\bm x}-\frac{1}{2}\right)-\frac{\gamma}{4}  ,
\ee
where $c_{\bm x}$ and $d_{\bm x}$ ($c^\dagger_{\bm x}$ and $d^\dagger_{\bm x}$) are annihilation (creation) operators acting on the left- and right-kets in the doubled Hilbert space.
$w$ is the strength of hopping temrs, and $\hat{i}$ [$i=x,y$] represents the unit vector along the $i$ direction.

Now it is clear that the fidelity $P(t)=\tr \;e^{\cL t}$ is nothing but the canonical partition function of the non-Hermitian quantum system  $Z=\tr\;e^{-\beta H_{\rm eff}}$, with the Hamiltonian $H_{\rm eff}=-\cL$, and the inverse temperature $\beta=t$.
The left- and right-kets in the doubled Hilbert space is understood as the spin degrees of freedom, 
and the effective non-Hermitian Hamiltonian is the attractive Hubbard model with the non-Hermitian imaginary hopping terms.
We note that we can absorb the imaginary unit $i$ in Eq.~\eqref{eq:hopping} into the definition of fields: $c^\prime_{\bm x}=i c_{\bm x}$, and $d^\prime_{\bm x}=i d_{\bm x}$.
By this redefinition, Eq.~\eqref{eq:hopping} is equivalent to the Hatano-Nelson type hopping terms discussed in Ref.~\cite{Hayata:2021erf}.

By using the Suzuki-Trotter decomposition 
\be
e^{\cL t}=\Pi_{n=1}^{N_t} e^{K \Delta t}e^{U \Delta t} +{\rm O}(\Delta t^2),
\ee
with $\Delta t=t/N_t$, and $N_t$ being the number of the steps in the Suzuki-Trotter decomposition, 
and the Hubbard-Stratonovich transformation~\cite{PhysRevB.28.4059} 
\be
e^{\Delta t\gamma  \left(c^\dagger_{\bm x} c_{\bm x}-\frac{1}{2}\right) \left( d^\dagger_{\bm x} d_{\bm x}-\frac{1}{2}\right)}
=\frac{e^{-\frac{\gamma\Delta t}{4}}}{2}\sum_{s_{\bm x}=\pm1}e^{s_{\bm x}\lambda(c^\dagger_{\bm x} c_{\bm x}+d^\dagger_{\bm x} d_{\bm x}-1)} ,
\ee
with $\cosh\lambda=e^{\gamma \Delta t/2}$, the fidelities in Eqs.\eqref{eq:echo}, and~\eqref{eq:purity} are written as~\cite{Blankenbecler:1981jt,PhysRevB.40.506}
\be
\begin{split}
M(t,\lambda)  &=\tr\; e^{-K t}\Pi_{n=1}^{N_t} e^{K \Delta t}e^{U\Delta t} 
\\
&={\cal N}\sum_{s_{n,\bm x}=\pm1}
e^{-\lambda s_{n,\bm x}}{\rm det}\left[1+e^{-K_\up t}B_\up\right]
\\
&\;\;\;\;\times{\rm det}\left[1+e^{-K_\down t}B_\down\right] 
\\
&={\cal N}\sum_{s_{n,\bm x}=\pm1}e^{-\lambda s_{n,\bm x}}\left|{\rm det}\left[1+e^{-K_\up t}B_\up\right]\right|^2 
\\
&\eqqcolon  \sum_{s_{n,\bm x}} m(t,\lambda) ,
\end{split}
\ee
and
\be
\begin{split}
P(t,w) &=\tr \; \Pi_{n=1}^{N_t} e^{K \Delta t}e^{U \Delta t}
\\
&={\cal N}\sum_{s_{n,\bm x}=\pm1}
e^{-\lambda s_{n,\bm x}}{\rm det}\left[1+B_\up\right]
\\
&\;\;\;\;\times{\rm det}\left[1+B_\down\right] 
\\
&={\cal N}\sum_{s_{n,\bm x}=\pm1}e^{-\lambda s_{n,\bm x}}\left|{\rm det}\left[1+B_\up\right]\right|^2 
\\
&\eqqcolon  \sum_{s_{n,\bm x}} p(t,w) ,
\end{split}
\ee
where $K_{\sigma=\up,\down}$ are the hopping matrices with amplitude $\mp iw$, and $B_{\sigma=\up,\down}=e^{K_\sigma\Delta t}e^{\lambda V(s_{1,\bm x})}\cdots e^{K_\sigma\Delta t}e^{\lambda V(s_{N,\bm x})}$, with $V(s_{N,\bm x})$ being the diagonal matrix, whose components are $s_{n,\bm x}=\pm1$.
 The normalization factor is ${\cal N}=\left(e^{-\frac{\gamma\Delta t}{2}}/2\right)^{N_tV}$, with $V$ being the spatial volume.
Since $e^{K_\down \Delta t}=\left[e^{K_\up \Delta t}\right]^*$, the weights $m(t,\lambda)$, and $p(t,w)$ are semi-positive, so that we can compute  the time evolution of the fidelities based on the quantum Monte Carlo.
Importantly, the Hamiltonian dynamics evolves the two spin components along the forward and backward directions, so that the complex phase is always cancelled in the spin components, while the weight of each spin component is complex and oscillates, which strongly harms the importance sampling only with the one-sided time evolution. 

In Monte Carlo simulations, we cannot evaluate  the weight itself. 
As is common in them, we regard the ratio of weights as an observable and compute $M(t,\lambda)$ as 
\be
\begin{split}
\frac{M(t,\lambda)}{M(t,0)} &=\prod_{i=0}^{N-1}\frac{M(t,(i+1)\Delta\lambda)}{M(t,i\Delta\lambda)}
\\
&=\prod_{i=0}^{N-1}\frac{1}{\langle\frac{m(t,i\Delta\lambda)}{m(t,(i+1)\Delta\lambda)}\rangle_{t,(i+1)\Delta\lambda}} ,
\end{split}
\label{eq:echo_ratio}
\ee
where $M(t,0)=\tr\bm1=2^{2V}$, $\Delta\lambda=\lambda/N$, and 
\be
\langle O(s_{n,\bm x})\rangle_{t,i\Delta\lambda}=\frac{\sum_{s_{n,\bm x}} m(t,i\Delta\lambda)O(s_{n,\bm x})}{\sum_{s_{n,\bm x}} m(t,i\Delta\lambda)} .
\ee
We note that the ratio of the normalization factor should be correctly taken into account in Eq.\eqref{eq:echo_ratio}.
Instead of computing $M(t,\lambda)/M(t,0)$ itself, we rewrite $M(t,\lambda)/M(t,0)$ as the product of the ratios as in Eq.\eqref{eq:echo_ratio}, and compute each ratio indepedently by the quantum Monte Carlo.
In this way, each ratio can be computed efficiently since the typical set of ${s_{n,\bm x}}$ has a good overlap in the two fidelities.
Similarly, $P(t,w)$ is computed as
\be
\begin{split}
\frac{P(t,w)}{P(t,0)} &=\prod_{i=0}^{N-1}\frac{P(t,(i+1)\Delta w)}{P(t,i\Delta w)}
\\
&=\prod_{i=0}^{N-1}\frac{1}{\langle\frac{p(t,i\Delta w)}{p(t,(i+1)\Delta w)}\rangle_{t,(i+1)\Delta w}}
\end{split}
\label{eq:purity_ratio}
\ee
where $P(t,0)=[2(1+e^{-\gamma t})]^V$ can be analytically computed, $\Delta w=w/N$, and
\be
\langle O(s_{n,\bm x})\rangle_{t,i\Delta w}=\frac{\sum_{s_{n,\bm x}} p(t,i\Delta w)O(s_{n,\bm x})}{\sum_{s_{n,\bm x}} p(t,i\Delta w)} .
\ee

\begin{figure}[t]
\begin{center}
 \includegraphics[width=.48\textwidth]{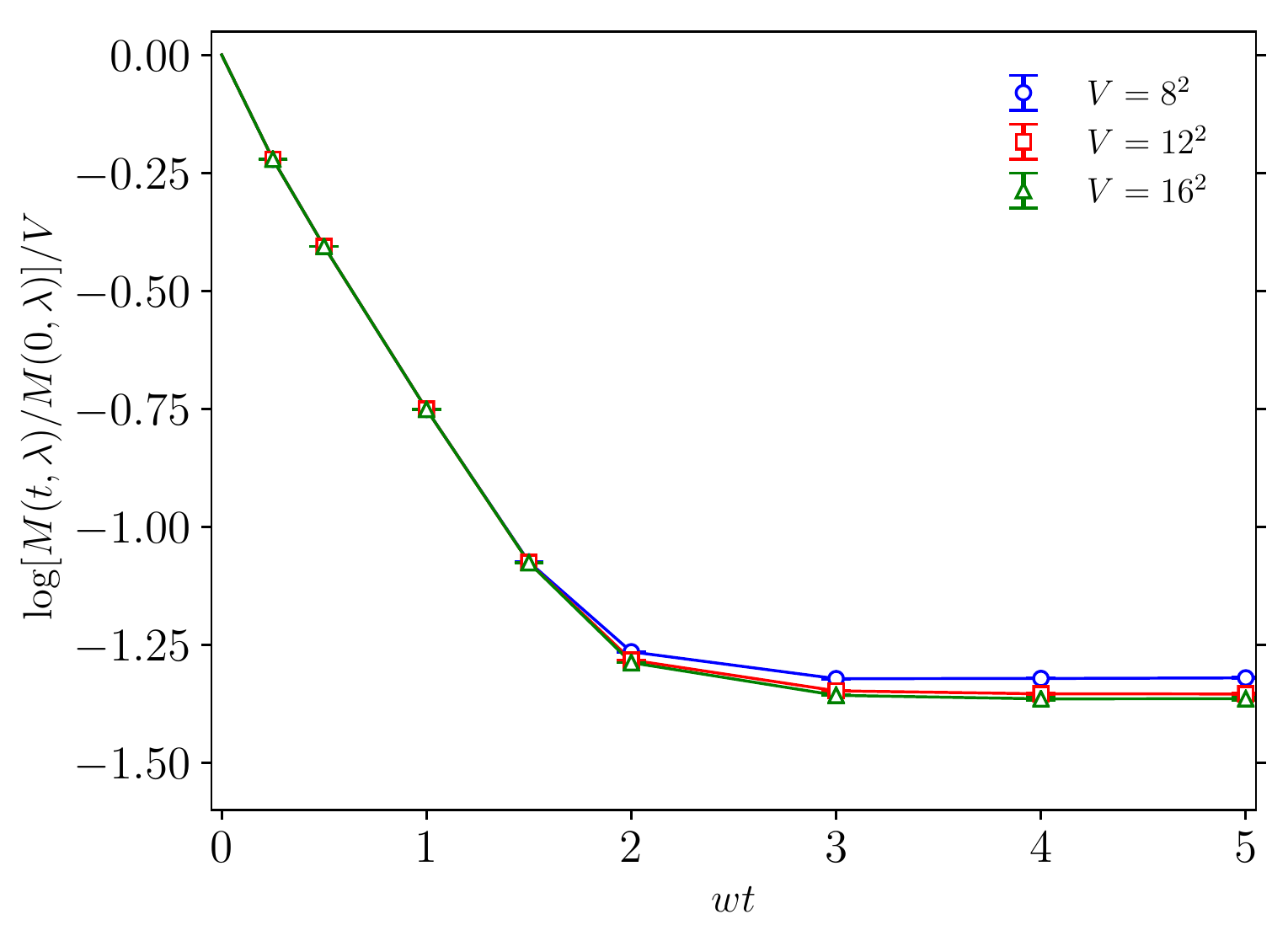}
\caption{\label{fig:echo}
Time evolution of the generalization of Loschmidt echo $M(t,\lambda)$. We normalize $M(t,\lambda)$ by its initial value $M(0,\lambda)=2^{2V}$. The strength of the coupling is $\gamma/w=4$. The error bars are the standard error of the mean.
}
\end{center}
\end{figure}

\paragraph{Numerical simulation.} 
We computed the ratios of the fidelities in Eqs~\eqref{eq:echo_ratio}, and~\eqref{eq:purity_ratio} by the the determinant quantum Monte Carlo~\cite{Blankenbecler:1981jt}.
The simulation details are as follows.
We take the strength of quantum jumps $\gamma/w=0.1$ and $4$, and the total number of the products $N=32$.
We performed the simulations with one lattice volume $V=8^2$ for $\gamma/w=0.1$, and with three lattice volumes $V=8^2$, $12^2$, and $16^2$ for $\gamma/w=4$.
We fixed the Trotter step with $\Delta t=0.05/w$.
The extrapolation $\Delta t\rightarrow0$ is left for a future study. 
We employed the stabilization techniques developed in Refs.~\cite{PhysRevB.40.506,BAI2011659} to compute a long Trotter chain without deteriorating the numerical precision.

\begin{figure}[t]
\begin{center}
 \includegraphics[width=.48\textwidth]{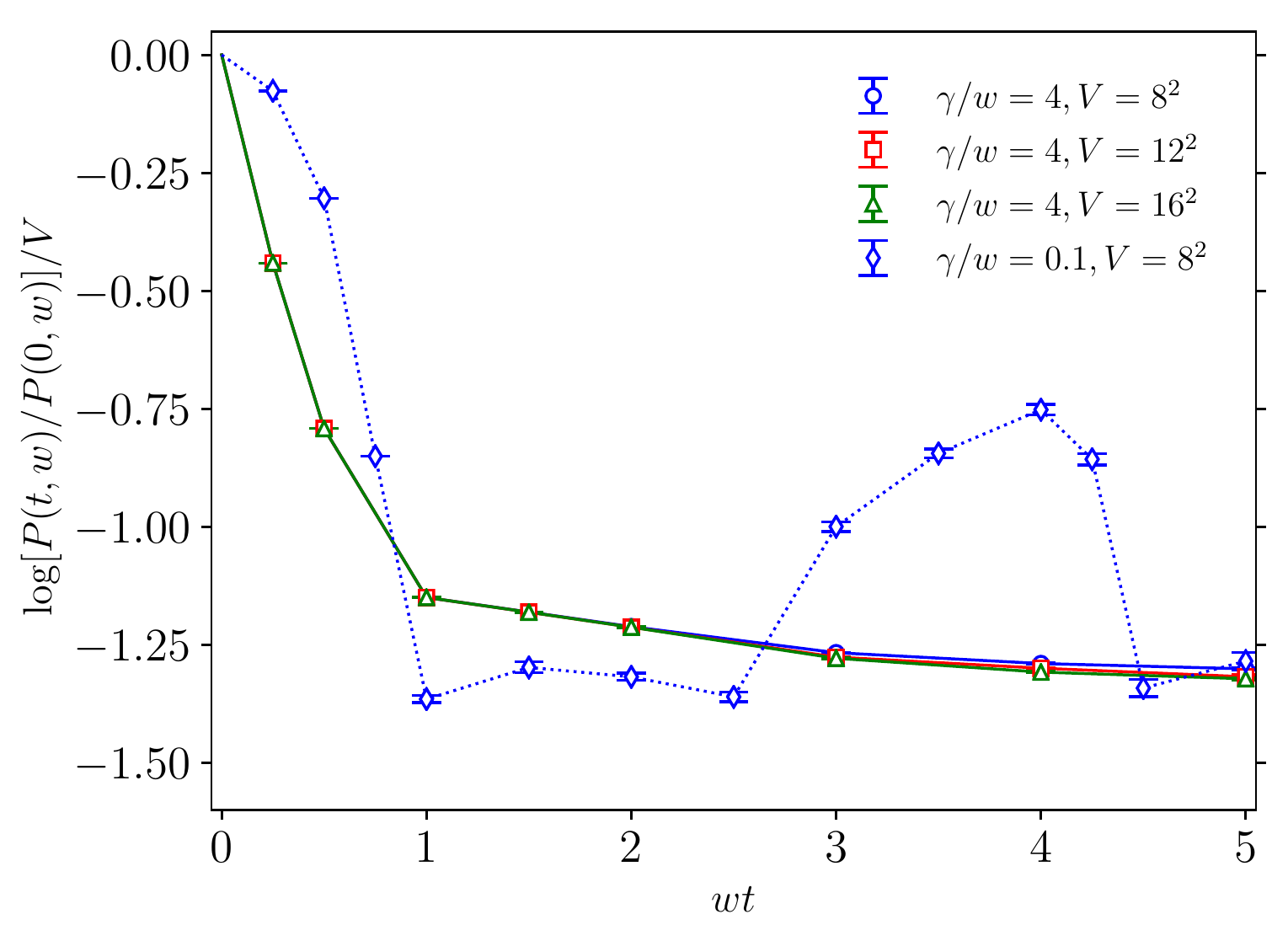}
\caption{\label{fig:purity}
Time evolution of the relative purity $P(t,w)$. 
We normalize $P(t,w)$ by its initial value $P(0,w)=2^{2V}$.
The solid curves show the results with $\gamma/w=4$, and $V=8^2$, $12^2$, $16^2$, while the dashed curve shows the results with $\gamma/w=0.1$, and $V=8^2$.
The error bars are the standard error of the mean.
}
\end{center}
\end{figure}

We show the time evolution of  the generalization of Loschmidt echo $M(t,\lambda)$, and the relative purity $P(t,w)$ in Figs.~\ref{fig:echo}, and~\ref{fig:purity}.
For $\gamma/w=4$, we clearly see that both of $M(t,\lambda)$ and $P(t,w)$ decay exponentially in time, and reach the stationary values very quickly.
We found that the decay rates and stationary values show the volume-law scaling as clearly seen in the volume dependence of the fidelities.

We have also performed numerical simulations with the small coupling strength $\gamma/w=0.1$.
We show the time evolution of  the relative purity $P(t,w)$ in Fig~\ref{fig:purity}.
Surprisingly, using the quantum Monte Carlo, we can compute even the oscillatory damping.  
This oscillation originates from the unitary time-evolution of free fermions, so that it is cancelled in $e^{-\cL_{\gamma=0}t}$ and $e^{\cL t}$, and not visible in $M(t,\lambda)$.

 \paragraph{Summary and future prospects.} 
 We have shown that the real-time evolution of fidelities in open fermion systems can be computed on the basis of the determinant quantum Monte Carlo.
To this end, we prescribe a mapping between the fidelities in open quantum systems and the partition function in non-Hermitian quantum systems; the latter can be solved with the determinant quantum Monte Carlo.
Although we considered the simple spinless fermion model for a demonstration, 
the common strategy to search the sign-free Hamiltonians in Hermitian systems is still useful for the non-Hermitian Hamiltonian~\cite{Hayata:2021erf}, and thus we can find more generic sign-free open fermion systems.

There are several future applications.
First, although we consider the averaged fidelities to map a real-time open quantum system to a finite-temperature non-Hermitian quantum system (in particular for identifying time as inverse temperature), we can also start with a specific initial state without averaging.
In this case, the fidelities in Eqs.~\eqref{eq:echo_pro}, and~\eqref{eq:purity_pro} can be computed on the basis of the projector quantum Monte Carlo.
This might be more suitable for a comparison with experiments.
Second, we may be able to compute the fidelity-type out-of-time-order correlators~\cite{PhysRevA.94.040302,G_rttner_2017} by generalizing our method.
This would be useful to study the information scrambling under dissipation.
Finally, we have computed the fidelities of an open quantum system, or equivalently, the partition function of a non-Hermitian quantum system. 
We can compute correlation functions as in the usual quantum Monte Carlo simulations.
In particular, our method can be used to study the finite-temperature phases of non-Hermitian quantum systems via the mapping to the real-time open quantum systems.
This would be very interesting since the physical meaning of temperature was not clear in non-Hermitian quantum systems,  and only the phases of the groundstate have been discussed so far.

 \acknowledgments
 
 This work was supported by JSPS KAKENHI Grant Numbers 21H01007, and 21H01084.
 The numerical calculations were carried out on Yukawa-21 at YITP in Kyoto University, and on cluster computers at iTHEMS in RIKEN.

\bibliographystyle{apsrev4-2}
\bibliography{fidelity}

\end{document}